\documentclass[aps,prl,twocolumn,superscriptaddress]{revtex4-1}
\usepackage{graphicx}
\usepackage{amsmath}
\graphicspath{{./Bilder/Bilder_pdf/}}

\begin{document}

\preprint{}

\title{Phase transition for parameter learning of Hidden Markov Models}


\author{Nikita Rau}
\affiliation{Institut f\"ur Physik, Universit\"at Oldenburg, D-26111 Oldenburg, Germany}
\author{J\"org L\"ucke}
\affiliation{Department of Medical Physics and Acoustics, Universit\"at Oldenburg, D-26111 Oldenburg, Germany}
\author{Alexander K. Hartmann}
\affiliation{Institut f\"ur Physik, Universit\"at Oldenburg, D-26111 Oldenburg, Germany}


\date{\today}

\begin{abstract}
We study a phase transition in parameter learning of  Hidden Markov Models
(HMMs).
We do this by generating sequences of observed symbols from given
discrete 
HMMs with uniformly distributed transition probabilities and a noise level
encoded in the output probabilities. 
By using the Baum-Welch (BW) algorithm, an Expectation-Maximization 
algorithm from the field of Machine Learning, we then try to estimate
the parameters of each investigated realization of an HMM. 
We study HMMs with $n=4, 8$ and 16 states.
By changing the amount of accessible learning data and the noise level, 
we observe a phase-transition-like change in the performance of the 
learning algorithm. For bigger HMMs and more learning data, the 
learning behavior
improves tremendously below a certain threshold in the noise strength.
For a noise level above the threshold, learning is not possible.
Furthermore, we use an overlap parameter applied to the
results of a maximum-a-posteriori (Viterbi) algorithm to investigate 
the accuracy of the hidden state estimation around the phase transition.
\end{abstract}


\maketitle

\section{Introduction}

Phase transitions \cite{stanley1971,yeomans2000} 
are phenomena of central interest in physics and in 
particular statistical physics and thermodynamics. Classically,
 phase transitions are studied for
actual physical system like liquid-vapor transitions of gases,
ferromagnetic transitions of magnets or the 
super conduction phase transition of metals. The behavior of
phase transitions becomes more interesting if (quenched) 
disordered systems are studied,
for example for the percolation transition,
the spin glass-paramagnet transition of spin glasses or
the localization transition of disordered Bose systems. Since some decades,
also phase transitions in ``non-physical'' systems are studied, e.g.,
the jamming transition in transport models like
the Nagel-Schreckenberg model \cite{nagel1992}, the transition to an
epidemic state in disease spreading \cite{newman2002}, ``easy-hard''
phase transitions in optimization problems \cite{phase-transitions2005} 
or the transition to synchronicity of 
brain activity as described by the Kuramoto model \cite{kitzbichler2009}. Also 
information-theoretic phase transitions with respect to analyzing
(large) sets of data have become a field of interest, e.g.,
when finding communities in 
networks \cite{newman2004communityA,ronhovede2010,decelle2011,hu2012}
analyzing the complexity of data generated by random systems 
\cite{entropy_complexity_random2013},  
learning of patterns in neural networks \cite{seung1992} 
and detecting causality in
Bayesian networks \cite{causality_pairs_triplets}.
Many of these information-theoretic phase transitions seek to 
distinguish between
phases where the desired information, like the structure of communities
or the direction of causal interactions, can be obtained from the
given data, and for phases where this is not possible. 
Investigating these phase
transitions allows one to understand the fundamental limitations
of learning and extracting information from the data in general
and in dependence of the used models and algorithms.
This is a fundamental way to look at many problems and approaches
 which are considered in the field of \emph{machine 
learning} \cite{bishop2006,faul2020}, which
has become in recent years of major interest not only for ubiquitous 
applications but also for fundamental scientific studies. 
Note that, interestingly, machine-learning
models like neural networks have been used also as tools to extract
phase transitions in different system like Ising systems 
\cite{wetzel2017,carrasquilla2017,kashiwa2019}.

Nevertheless, in this work we are interested in the first mentioned connection
between phase transitions and data analysis, i.e., the question whether
there exist a transition between a phase where the fundamental
parameters of a model can be extracted from the given data, and a phase
where not.
Specifically, we study the behavior of learning of parameters of 
elementary Hidden Markov Models (HMMs) \cite{rabiner1989,durbin1998}
by computer simulations \cite{practical_guide2015}. HMMs are widespread
in data analysis and modeling, e.g., speech-recognition \cite{rabiner1989},
biological sequence analysis \cite{won2004}, or analysis of
gestures \cite{wilson2001}. 

Although HMMs have been used often as tools, also in physical contexts,
e.g., to treat data in experimental physics \cite{kanter2005}
or analyze phase transitions in physical systems \cite{bechhoefer2015},
they have, to our knowledge, only rarely been the
object of interest in a physical study, in particular with respect to
phase transitions occurring in the HMMs.
 For example, the entropy of a binary HMM was
calculated \cite{zuk2005} by a mapping to a one-dimensional Ising model. 
Lathouwers and Bechhoefer have investigated  
\cite{lathouwers2017} transitions
with respect to whether the reconstruction of a hidden sequence is possible
or not depending on whether data can be kept in memory. Allahverdyan and
Galastyan have investigated \cite{allahverdyan2009,allahverdyan2015} 
the maximum a posteriori (Viterbi)
 sequence as a function of a noise parameter and found
transitions between regions where almost full sequence reconstruction
is possible and regions where not.

In contrast to these previous works, as mentioned, we are
not interested in analyzing the properties or 
performance of a given HMM, with known parameters, with respect
to the given data, but we are interested whether it is possible
to learn  the unknown parameters of a HMM from the given data. 
Specifically we will numerically generate data for a HMM with given ``ground truth'' 
parameter set, where we control some noise via the emission probabilities and
subsequently try to learn the parameters again using the 
BW algorithm \cite{baum1966,durbin1998}. We
analyze the learning of the transition and emission
probabilities specifying a HMM.
We
are interested whether there exist a sharp transition between a ``learning
phase'' and a phase where the determination of the parameters fails.
As we will see below, this is indeed the case.

The reminder of the paper is organized as follows: First, we present
the definition of an HMM and state the ensemble of random HMMs we have used.
Next, we explain the algorithms we applied to simulate HMMs, to
calculate posterior probabilities and to learn the parameters from the data.
In the following chapter, we define the measurable quantities we have
recorded. In the main part, we then present our simulation results.
We finish by a summary and discussion.

\section{Definitions}

Here we present the definitions we use in the present work, in particular
of the Hidden Markov  Model.
HMMs consist of a finite set of $n$ (hidden) states 
 and a finite or infinite 
 set of emission symbols. A chain generated by a HMM starts in some initial state,
which is randomly chosen with probabilities given by 
a vector $\overrightarrow{A^0}=(A^0_1,\ldots A^0_n)$. Transitions between
states $i,j$ occur at discrete steps 
with probabilities $A_{ij}$, which
is the probability to go into state $j$ in the next step if the HMM is
in state $i$ in the current step.
 These probabilities
are collected in an $n \times n$ transition
matrix $\textit{\textbf{A}}$. Since the probability to be
in a certain state depends only on the previous state, the sequence of states,
denoted by $\overrightarrow{x}=(x_1,.\ldots,x_L)$ ($L$: length of sequence),
forms a Markov chain. Nevertheless, the states are hidden, i.e., cannot
be observed. Instead, at each state a randomly draw symbol is emitted,
creating a sequence $\overrightarrow{y}=(y_1,.\ldots,y_L)$. Here
we consider the discrete case where each time one symbol from an $m$ letter
alphabet is emitted. Let $B_{ik}$ denote the probability to
emit symbol $k$ in state $i$. These probabilities are collected
in the $n\times m$ matrix $\textit{\textbf{B}}$.
The regular
conditions for probabilities apply, 
i.e. all entries are non-negative and the entries are
normalized:

\begin{equation}
\begin{aligned}
&\sum_{i=1}^{n} A^0_{i}=1;
\\&\sum_{j=1}^{n} A_{ij}=1 \text{ } \forall i \in \{1,...,n\} ;
\\&\sum_{k=1}^{m} B_{ik}=1  \text{ } \forall i \in \{1,...,n\}
\end{aligned}
\label{eq:normalization}
\end{equation}
Here and in the following, we will always use letter $i,j$ to indicate states 
and the letter $k$ to indicate an emission symbol.
In summary, each HMM is characterized by the set of parameters
$\theta=(\overrightarrow{A^0},\textit{\textbf{A}},\textit{\textbf{B}} )$.  
In this work, the HMMs are
chosen in a way that there are as many emission symbols as states, i.e., $m=n$. 
Furthermore, $\textit{\textbf{A}}$ and $\textit{\textbf{B}}$ are chosen 
to have a  specific structure. For the transition matrices, we consider an ensemble of quenched
disorder matrices which all  have the form 

\begin{equation}
A_{ij} =
\begin{cases}
p_T & \text{ if } i=j \\
\frac{1-p_T}{n-1} & \text{ otherwise }
\end{cases} \,.
\label{eq:structure_transition}
\end{equation}
For each disorder realization matrix $p_T$ is a uniformly distributed value
drawn from interval $[0.85,1]$. Thus, for each matrix 
$p_T$ is the probability for remaining in the current
hidden state. For  big values of $p_T$, the transition into other
hidden states different from the  current state is less probable, i.e., the
hidden state chain will exhibit less fluctuations.
We have restricted the values to  $p_\ge0.85$ to reduce the fluctuations 
which makes our simulations less demanding in terms of statistics.
 The general result would not change if we allow for a large range of 
$p_T$ as we have verified for some test cases.

Furthermore, for each disorder realization the vector of initial-state probabilities
consist of a set of $U(0,1)$ uniformly drawn random numbers, such that the sum of these
number is normalized to one, i.e.,

\begin{equation}
\label{eq:initmatrices}
A^0_i = r_i/\sum_{i=1}^n r_i\quad \text{with } r_i \sim U(0,1)\,
\end{equation}

The emission matrices have the form
\begin{equation}
B_{ik} =
\begin{cases}
p_{\rm E} & \text{ if } i=k \\
\frac{1-p_{\rm E}}{m-1} & \text{ otherwise }
\end{cases}\,,
\label{eq:structure_emission}
\end{equation} 
where 
  $p_{\rm E}\in [1/m,1]$ is a fixed (external) parameter that controls the
 output noise level of the HMM: The case when $p_{\rm E}$ has
the value $1$ corresponds to a HMM with no noise at all:
In this case each emission symbol of a hidden state corresponds
to the hidden state
itself and all other symbols are not emitted. The lower bound $p_{\rm E}=\frac{1}{m}$
represents a HMM with a maximum noise level. In this case columns of
$\textit{\textbf{B}}$ are all the same, therefore the hidden states
can not be distinguished by their emission probabilities. 

For each given HMM, we generate Markov chains of states and corresponding
sequences of emitted symbols. Note that for each HMM, and correspondingly
each sequence, the parameters actually used are called \emph{ground truth}
parameters. The aim of our work is to see how well
the learned parameters agree with the ground truth, see below for our measurable
quantities.  
All results will be averages over a suitable number of matrices
drawn from this ensemble. Note 
that the ensemble is a generalization and extension of the non-disordered, i.e., 
fixed matrices which were considered previously  
for the smallest possible case of  $n=2$  states  \cite{allahverdyan2009}.

For all our work we consider different values of $p_{\rm E}$, but all averages
over different transition matrices $\textit{\textbf{A}}$
will be performed for fixed values of $p_{\rm E}$, i.e., the same
matrix $\textit{\textbf{B}}$. All results are then analyzed
as a function of the parameter value $p_{\rm E}$. We expect
that in the limit $p_{\rm E}\to 1$ it will be much easier for any algorithm
to learn the parameters from the sequence of visible symbols,
while for the limit $p_{\rm E}\to\frac{1}{m}$ it will become impossible.
In particular we are interested in whether between these limiting values
there exist a transition from an ``easy'' learning phase, at large
values of $p_{\rm E}$ to a ``hard''
learning phase for small values of $p_{\rm E}$.

\section{Algorithms}
In this work, the parameter learning is executed by the Baum-Welch 
algorithm \cite{baum1966}. The algorithm is an Expectation-Maximization algorithm
which seeks parameters $\theta^*$ that maximize the data likelihood $P(\overrightarrow{y}^1, 
..., \overrightarrow{y}^N|\theta)$ of a given training data set 
$\{\overrightarrow{y}^1, \ldots, \overrightarrow{y}^N\}$, i.e.: 
\begin{equation}
\theta^*= \text{ argmax}_\theta P(\overrightarrow{y}^1, ..., \overrightarrow{y}^N|\theta)\,.
\label{eq:likelihood}
\end{equation} 

Given an HMM model and initial parameters $\Theta$, the training data set can be considered
the input to the algorithm, and the parameters $\Theta^*$ can be considered the output.
The BW algorithm is like EM algorithms in general an iterative procedure that can converge
to local optima of the data likelihood. The BW algorithm is very standard choice and preferable,
e.g., to Viterbi training \cite{durbin1998} which does in general not converge to (possibly local)
maxima of the data likelihood.
%
%
 
For comprehensiveness, we outline
the algorithm here, details can be found in the literature.  One starts
with first-guess initializations 
$\theta=(\overrightarrow{A^0},\textit{\textbf{A}},\textit{\textbf{B}} )$, 
which are uniformly drawn here. 
In each iteration, the algorithm
proceeds in two steps which will be presented in more detail in the
following sections: In the expectation step, the E-step, the
BW algorithm calculates the expected times of transitions between two
hidden states, the expected times of symbol emissions by hidden states
and the expected number of times a sequence starts with a certain
hidden state. Based on these calculations in the maximization step,
the M-step, the new parameters are calculated. The algorithm
guarantees a step-wise decrease of the Kullback-Leibler distance
between the probability distributions over symbol sequences of the
data and the model \cite{breuer1996}, i.e., the data likelihood
increases monotonously.

\subsection{E-step}

This step requires sum over all hidden paths $\overrightarrow{x}=(x_1,.\ldots,x_L)$ 
which are compatible with one given  observation $\overrightarrow{y}$ (taken
from $\overrightarrow{y}^1, ..., \overrightarrow{y}^N$).
 For this purpose so-called 
forward-variables $f_i(l)$ and backward-variables $b_i(l)$ in Eq. \ref{eq:forward_backward}
are calculated:

\begin{equation}
\begin{aligned}
f_i(l)&= P(y_1,...,y_l, x_l=i), \text{ with } l \in \{1,...,L\}
\\b_i(l)&=P(y_{l+1},...,y_L|x_l=i), \text{ with } l \in \{1,...,L-1\}\,.
\end{aligned}
\label{eq:forward_backward}
\end{equation}

The forward variable $f_i(l)$ describes the joint probability that the
hidden state $i$ occurs at the $l$-th position of a sequence and the
first $l$ observations were emitted. The backward variable $b_i(l)$ expresses
the conditional probability that the last $L-l$ observations occur 
conditioned on the $l$-th hidden state $x_l$ being $i$
\cite{durbin1998}. There are recursive calculation rules that enable one to get forward
and backward variables for every position within a sequence, see
\cite{rabiner1989}. Combining the sum rule for probabilities
$P(X,Y)=P(X|Y) \cdot P(Y)$ with the definitions of
$\textit{\textbf{A}}$, $\textit{\textbf{B}}$  and
Eq. \ref{eq:forward_backward}, one obtains \cite{durbin1998}

\begin{equation}
P(x_l=i, x_{l+1}=j| \overrightarrow{y}, \theta)=\frac{f_i(l) \cdot A_{ij} 
\cdot B_{jy_{l+1}} \cdot b_j(l+1)}{P(\overrightarrow{y})}
\label{eq:important_probability}
\end{equation}

Eq. \ref{eq:important_probability} represents the conditional
probability for getting the two   consecutive hidden states $i$ and $j$ at
the positions $l$ and $l+1$ under the condition that the whole
observation sequence $\overrightarrow{y}$ is
known. $P(\overrightarrow{y})$ can be calculated by using the
Forward-variables for $l=L$ as:
\begin{equation}
P(\overrightarrow{y})=\sum_{i=1}^{n}f_i(L)
\label{eq:probability_sequence}
\end{equation}

Using Eq. \ref{eq:important_probability} and Eq. \ref{eq:probability_sequence}
and by averaging over the data set
expected counts (denoted by an over bar) 
for the transition, emission and initial-state probabilities can
be obtained as

\begin{equation}
\overline{A_{ij}}=\sum_{n=1}^{N} \frac{1}{P(\overrightarrow{y}^n)} \sum_{l=1}^{L-1} f_i^n(l) 
\cdot A_{ij} \cdot B_{jy_{l+1}^n} \cdot b_j^n(l+1)
\label{eq:count_transition}
\end{equation}

\begin{equation}
\overline{B_{ik}}=
\sum_{n=1}^{N} \frac{1}{P(\overrightarrow{y}^n)} \sum_{\{l=1| y_l^n=k\}}^{L} f_i^n(l) \cdot b_i^n(l)
\label{eq:count_emission}
\end{equation}

\begin{equation}
\overline{A^0_{i}}=\sum_{n=1}^{N} \frac{1}{P(\overrightarrow{y}^n)} A^0_{i} 
\cdot B_{iy_{1}^n} \cdot b_i^n(1)
\label{eq:count_start}
\end{equation}

\subsection{M-step}

In the M-step, the parameters are updated to approach a (local) optimum. This is simply
achieved \cite{durbin1998}, similar to the case of maximum-likelihood
estimation for Gaussian distributions where the maximum-likelihood
parameter for the mean is the average. Here this is obtained   
by normalizing the expected counts Eqs.~\ref{eq:count_transition}- 
\ref{eq:count_start}:

\begin{equation}
A_{ij}^{\text{new}}=\frac{\overline{A_{ij}}}{\sum_{j'=1}^n \overline{A_{ij'}}}
\label{eq:new_transition}
\end{equation}

\begin{equation}
B_{ik}^{\text{new}}=\frac{\overline{B_{ik}}}{\sum_{i'=1}^m \overline{B_{i'k}}}
\label{eq:new_emission}
\end{equation}

\begin{equation}
(A^0_{i})^{\text{new}}=\frac{\overline{A^0_{i}}}{\sum_{i'=1}^{n} \overline{A^0_{i'}}}
\label{eq:new_start}
\end{equation}

E  and M steps are repeated until convergence. In this work for convergence we consider the
relative change of the data
likelihood $P(\overrightarrow{y}^1, ..., \overrightarrow{y}^N|
\theta)$ before and after the parameter update from
Eqs.~\ref{eq:new_transition}--\ref{eq:new_start}. When the relative
change is smaller than a threshold $\epsilon$, the BW
algorithm is terminated.

As we will see below, to which set of parameters the BW algorithm converges
depends on the choice of the initial parameter set. Therefore, as we will detail below,
we use BW with ten random restarts and select from the 10 outcomes the ``best''
one, i.e., that one with highest posterior probability.

\subsection{Viterbi algorithm}

For some of our simulations, we also computed the maximum a posteriori (MAP) path, i.e., the (hidden)
path $\overrightarrow{x^*}$ of states 
that maximizes for each observation $\overrightarrow{y}$ and given HMM parameters
 $\theta$ the path probability $P(\overrightarrow{x^*}|\overrightarrow{y}, \theta)$.
This can be done by the Viterbi algorithm \cite{Vit1967}. Similar to the 
forward-backward algorithm, it computes iteratively the Viterbi-variable $v_i(l)$. It is describing 
the probability of the most probable $l$-steps path  conditioned to it
ends in state $l$ and conditioned to the first $l$ letters of the observed sequence.
The hidden state sequence $\overrightarrow{x^*}$ can be obtained by backtracing.

\section{Setup, parameters and measurable quantities}\label{setup}
We applied the BW algorithm to ensembles of HMMs, as described by 
Eqs.~\ref{eq:structure_transition}--\ref{eq:structure_emission},
 for three different HMM sizes  $n \in \{4; 8; 16\}$.
 We have tested several values for 
the convergence parameter $\epsilon$.
We show results for $\epsilon= 10^{-7}$ because for higher values the convergence was a bit worse 
and for even smaller values the results do not
change substantially.
To see the influence of an increase of the available data, we have  
performed all numerical experiments for six different sizes $(N,L)$ of the learning
sets, for each HMM size, respectively.

Since the convergence of the BW algorithm depends on the initial parameter set, we have,
for each given realization of an HMM under consideration, 
run the BW algorithm 10 times with independently drawn initial
parameters, each resulting in a locally optimum estimate $\theta^*_r$ ($r=1,\ldots,10$).
  To select the best parameter set $\tilde \theta$ among the 10 
outcomes
of the 10 runs, 
we assume, to avoiding over-fitting effects, that  a second data set 
$\overrightarrow{z}^1, ..., \overrightarrow{z}^N$ of the same size is available
(or the available data was split into two halves).
The final best estimate $\theta'$ is that one which exhibits the maximum joint data probability 
$P(\overrightarrow{z}^1, ..., \overrightarrow{z}^N| \theta^*_r)$ $(r=1,\ldots,10)$.
For practical reasons, we consider $\log$ likelihood when possible, as usual. 
For technical convenience, when we add
up probabilities, we always normalize to some natural probability, as often done
\cite{durbin1998}.

All results presented below, for each considered value of $p_{\rm E}$, we have performed an average
over different realizations from the ensemble of HMMs. For $n=4$, we considered 1000
realizations, for $n=8$ we studied 600 realizations and for $n=16$ we found 200 realizations
to be sufficient.

Note that during the learning process it is assumed that the generating HMM-parameters and 
the hidden state sequences $\overrightarrow{x}^1, ..., \overrightarrow{x}^N$ of the 
learning data are unknown. Since we use artificially generated data they are nevertheless
available to us and we can use them as ground truth for comparison and 
evaluation of the learning process.
For our purposes, we measure the total error $E_{\rm tot}$ (Eq. \ref{eq:E_tot}), which is the sum of the absolute differences between
actual and estimated parameters:
\begin{equation}
\begin{aligned}
E_{\rm tot}&=\sum_{i=1}^{n} |A^0_{i}-
\tilde A^{0}_{i}| + \sum_{i=1}^{n} \sum_{j=1}^{n}|A_{ij}
-\tilde A_{ij}| \\
&+ \sum_{i=1}^{m}\sum_{k=1}^{n}|B_{ik}-\tilde B_{ik}|
\end{aligned}
\label{eq:E_tot}
\end{equation}
In general, also estimated parameters very different from the ground truth parameters can make up a successful run (e.g.\ in degenerated
cases when a model's likelihood is invariant under certain parameter permutations). In our case and for the way we choose the generating
parameters in Eqs. \ref{eq:structure_transition} to \ref{eq:structure_emission}, we found (\ref{eq:E_tot}) to measure the degree of
success of a given run sufficiently well.

Another way to test the estimated HMMs is to obtain, for each training sequence $\overrightarrow{y}^n$
 ($n=1,\ldots,N$) 
the most-likely hidden path $x_{1}^{*n}, \ldots x_{L}^{*n}$ by applying the
Viterbi algorithm to an HMM with the estimated parameters $\tilde\theta$. This can be compared to the 
actual paths $x_{1}^{i}, \ldots x_{L}^{i}$. The fraction of agreeing hidden states is given by the
so-called overlap $q$, which is a frequently used quantity 
in the physics of disordered systems:

\begin{equation}
q=\frac{\sum_{n=1}^{N} \sum_{l=1}^{L} \delta(x_{l}^{n}, x_{l}^{*n})}{N \cdot L}\,.
\label{eq:overlap}
\end{equation}
Here $\delta(x_{j}^{n}, x_{j}^{*n})=1$ if the hidden state of the learning set $x_{j}^{i}$ 
is equal to the hidden state of the Viterbi sequence $x_{j}^{*i}$ and otherwise zero. 
This means $q \in [0; 1]$ with $q=1$ corresponds to a $100\%$ reconstruction of the hidden 
sequences $\overrightarrow{x}^1, ..., \overrightarrow{x}^N$. Note that $q=1$ is not common
because even when using the true parameters of an HMM, the Viterbi path is only the most-likely
one, but very often not the actually generated one.

\section{Results}\label{results}

\begin{figure}
	\centering
	\includegraphics[width=0.99\columnwidth]{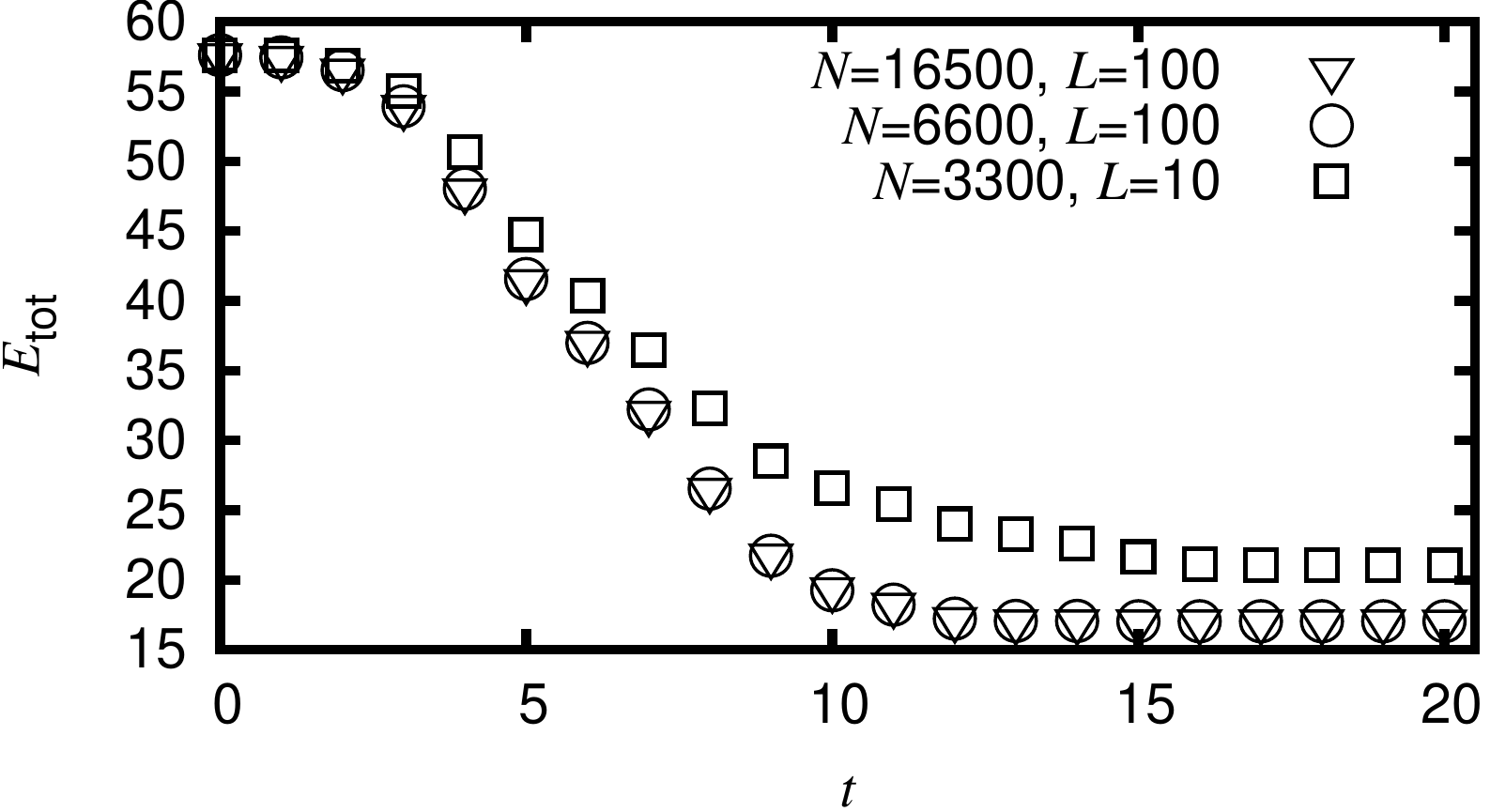}
	\caption{\label{fig:error_time}The evolution of the total error $E_{\rm tot}$ 
as a function of the 
BW step for a HMM with $n=m=16$ and three differently data set sizes, 
but the same initial parameters. }
\end{figure}

First, we study the behavior of the BW algorithm. In
Fig.~\ref{fig:error_time} the evolution of the total error $E_{\rm
  tot}$ is shown as a function of the step $t$ of the BW algorithm,
for three different learning data set sizes, but in all three cases
with the same set of starting parameters $\theta$ where each
parameter was drawn
uniformly form $[0,1]$.
Initially, the parameter set is very different $E_{\rm tot}\approx 57$
from the ground truth parameters of the original HMM, but during its
evolution, the error is decreased until it levels off at parameter
values still different from the ground truth ones, i.e., $E_{\rm tot}\gg 0$.  One also can
see that for larger data set sizes, the error is decreased more and
faster, but once a certain size is reached, no more improvement is
obtained. Below we will see that the combined size $NL$ acts like a
system size in the theory of phase transitions, which allows us to
extrapolate the phase transition point from ``hard'' to the
``easy'' learning phase.

\begin{figure}
	\centering
	\includegraphics[width=0.99\columnwidth]{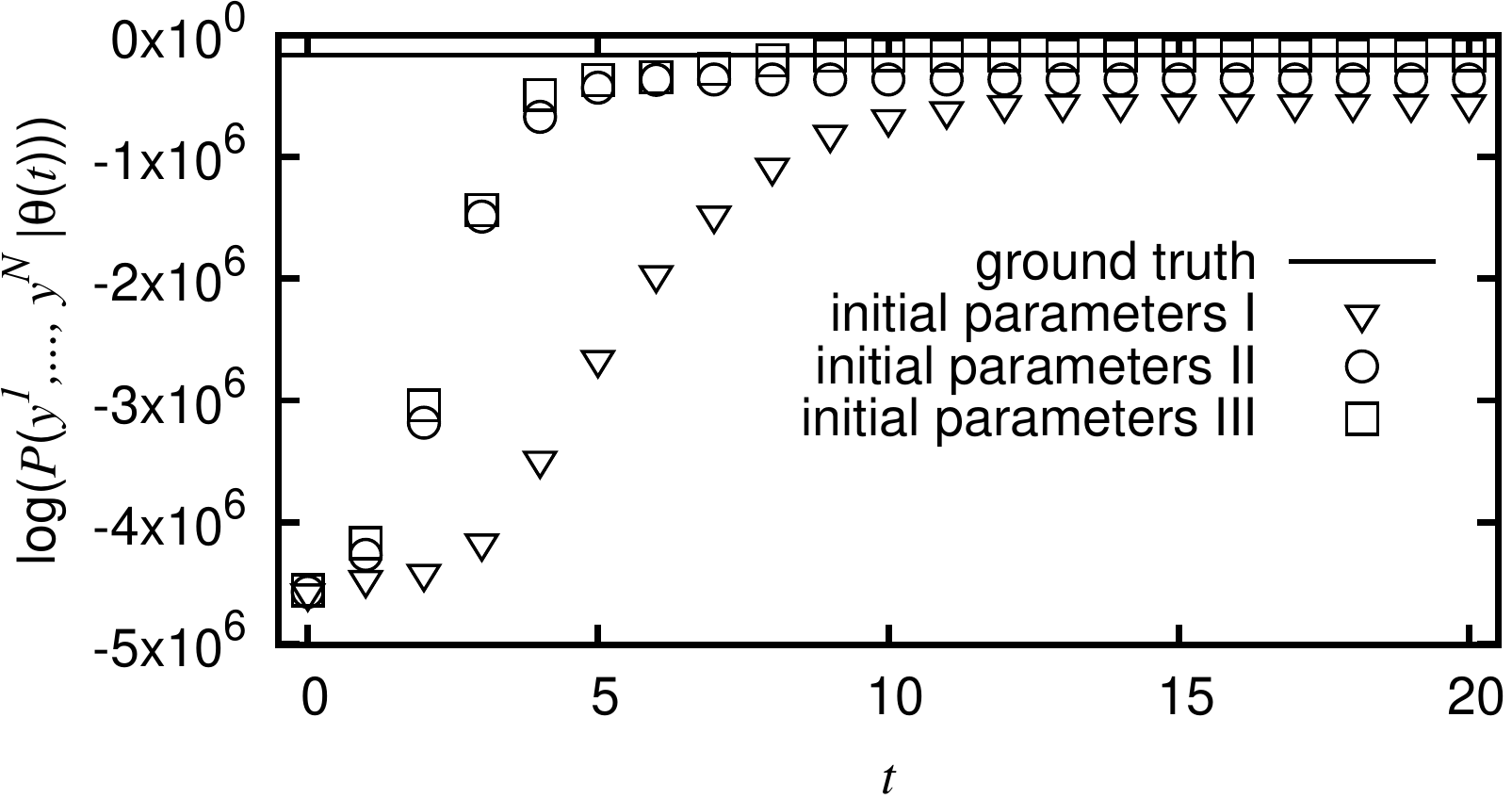}
	\caption{\label{ground_truth}The evolution of the log-likelihood as a function of the 
BW step for a HMM of size $n=m=16$ with three differently initial parameter sets 
$\theta^{\rm I}$, $\theta^{\rm II}$, and $\theta^{\rm III}$
as starting point of  the BW
algorithm. The log-likelihood was calculated for the learning set.
The solid line shows the ground truth. The corresponding total errors after
leveling off are: 
$E_{\rm tot, I}=17.005$, $E_{\rm tot, II}=4.362$ and $E_{\rm tot, III}=0.050$. 
}
\end{figure}

Since for the previous example the algorithm was not able to recover the
ground truth parameters, we
next study what influence the initial parameter set has. For three different, uniformly distributed, initial parameter sets
the log-likelihood $\log(P(\overrightarrow{y}^1, ...,
\overrightarrow{y}^N| \theta(t)))$ of the learning data for the current parameter set $\theta(t)$ is 
shown as function of the iteration $t$. One can observe in Fig.~\ref{ground_truth} that initially the growth in 
log-likelihood is fast, similar to the improvement seen for $E_{\rm tot}$ in Fig.~\ref{fig:error_time}.
After some iterations also these values level more or less off. One sees that
indeed for different initial parameters different final log-likelihoods are reached.
This illustrates the usefulness of repeated BW runs, from which the one with the highest
log-likelihood for the test data set is chosen and therefore the influence of
``unfortunate'' choices for the initial parameters is reduced.
Actually for  case (III) the log-likelihood 
is nearly equal to the ground truth, 
which was used to generate the data. This is an indication that indeed a very good estimate of the
parameters was obtained \cite{luecke2019}, which is supported by the fact that for this case
the error in parameters is $E_{\rm tot,III}=0.050$, i.e., very small.

\begin{figure}
	\centering
	\includegraphics[width=0.99\columnwidth]{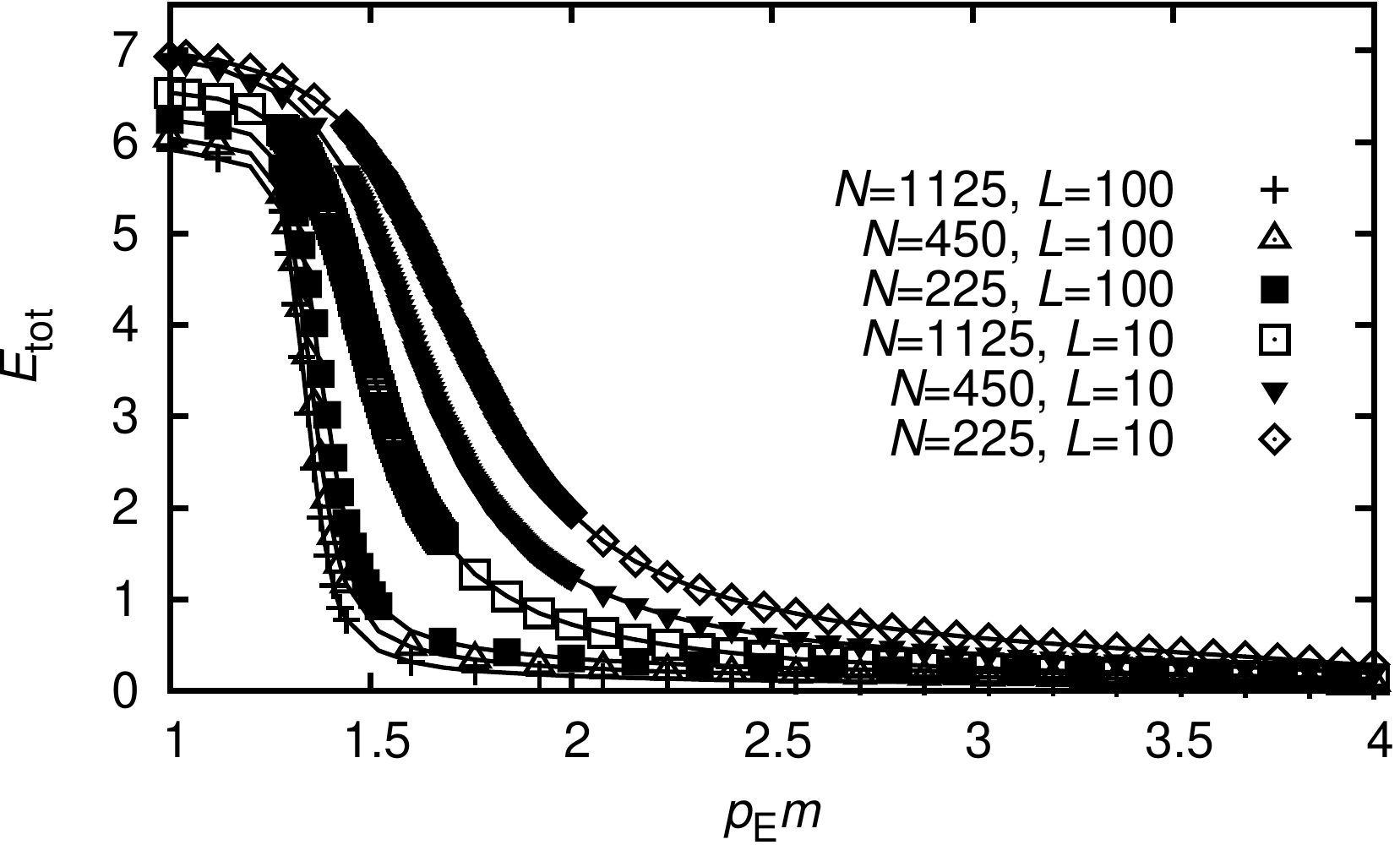}
	\caption{\label{tot_H4}The total error $E_{\rm tot}$ as a
          function of the noise parameter $p_{\rm E}$, multiplied by
          the number of symbols $m=4$ for six different learning set
          sizes.  Each data point is the average result over 1000
          simulations. The error bars are smaller than symbol size.
          For the three largest learning set sizes, the curves differ
          only slightly, which indicates that the results for the
          thermodynamic limit will look similar.}
\end{figure}

\begin{figure}
	\centering
	\includegraphics[width=0.99\columnwidth]{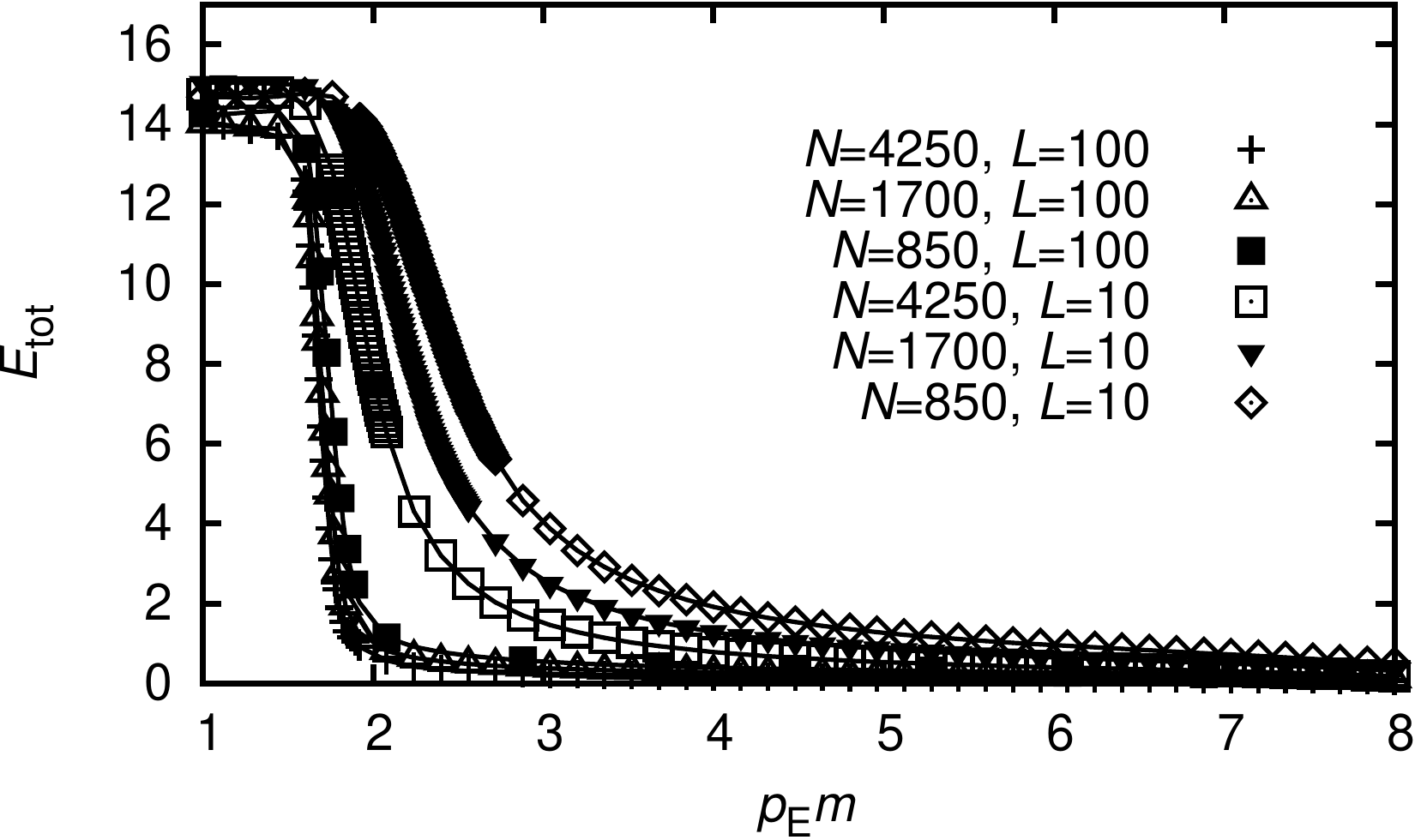}
	\caption{\label{tot_H8}The total error $E_{\rm tot}$ against the noise parameter $p_{\rm E}$, multiplied by the number of symbols $m=8$ for a better comparison of the different HMM sizes and for six different learning set sizes. Each data point is the average result of 600 simulations. The error bars are smaller than symbol size. For the three largest learning set sizes, the curves differ only slightly, which indicates that
the results for the thermodynamic limit will look similar.}
\end{figure}

\begin{figure}
	\centering
	\includegraphics[width=0.99\columnwidth]{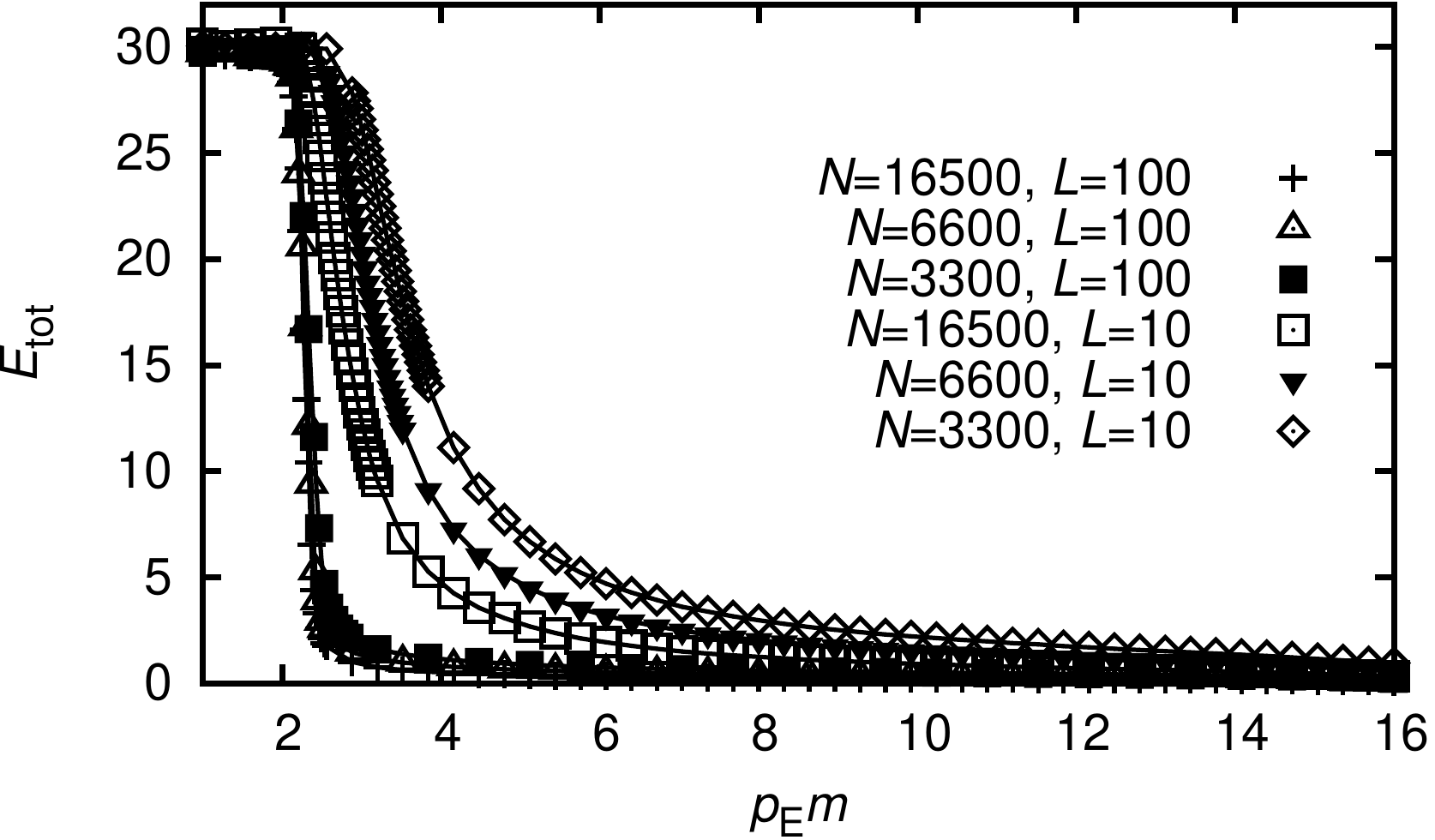}
	\caption{\label{tot_H16}The total error $E_{\rm tot}$ as a function of the noise parameter $p_{\rm E}$, multiplied by the number of symbols $m=16$ for six different learning set sizes. Each data point is the average result over 200 simulations. The error bars are smaller than symbol size.
For the three largest learning set sizes, the curves differ only slightly, which indicates that
the results for the thermodynamic limit will look similar.}
\end{figure}

In Fig. \ref{tot_H4} the total error $E_{\rm tot}$ is shown 
as a function of rescaled probability $p_{\rm E}m$ for
$m=n=4$ and six different learning set sizes $(N,L)$. It is
visible that for each learning set size the largest $E_{\rm tot}$-value,
i.e. the worst results, can
be found for $p_{\rm E} \cdot m=1 \leftrightarrow p_{\rm E}=0.25$ which meets the
expectations since $p_{\rm E}=0.25$ corresponds to the largest noise
level. For increasing $p_{\rm E} \cdot m$, the total error decreases and
reaches a minimum for $p_{\rm E} \to 1$ which corresponds to a
non-existing noise level,
where the full information about the states 
can be obtained from the data.  Hence, Fig. \ref{tot_H4} shows clearly that
the parameter-learning improves if the noise level is decreased.
Furthermore it is visible that the learning gets better 
by increasing the size of the
learning data set. The behavior for $N=1125$ and $L=100$ shows that
the biggest improvement occurs roughly in the interval $p_{\rm E} \cdot m
\in [1.3;1.4]$. The very steep decrease of the curve indicates a sharp change
from a non-learning to a learning behavior, i.e., a phase transition in the
information theoretic sense.
 A similar  drop of $E_{\rm tot}$ is observable for the
learning sets $N=450, L=100$ and $N=225, L=100$, which shows that the learning set
sizes are large enough to observe the limiting behavior.
The three
smallest learning sets show a less steeper decrease, indicating stronger finite size
effects.  It can also be
seen that the decline shifts to the left with increasing size of the
learning data set, which we will below use to determine a phase-transition point.

We have studied the behavior of the total error also for other HMM sizes.
Fig. \ref{tot_H8} and Fig. \ref{tot_H16}
show $E_{\rm tot}$ for $m=n=8$ and $m=n=16$. For each system size, the
parameter learning behaves qualitatively the same as for $n=m=4$. But one observes that 
in comparison to the case $n=m=4$ 
the decrease as a function of $p_{\rm E}m$ seems to become even steeper and its position
shifts slightly to larger parameter values, i.e., away from the point $p_{\rm E}m=1$ of
no information. This means, the size of the no-learning phase becomes bigger on the
rescaled $p_{\rm E}$ axis, indicating that for even larger HMM sizes the phase transition
from no learning to learning will still persist.

\begin{figure}
	\centering
	\includegraphics[width=0.99\columnwidth]{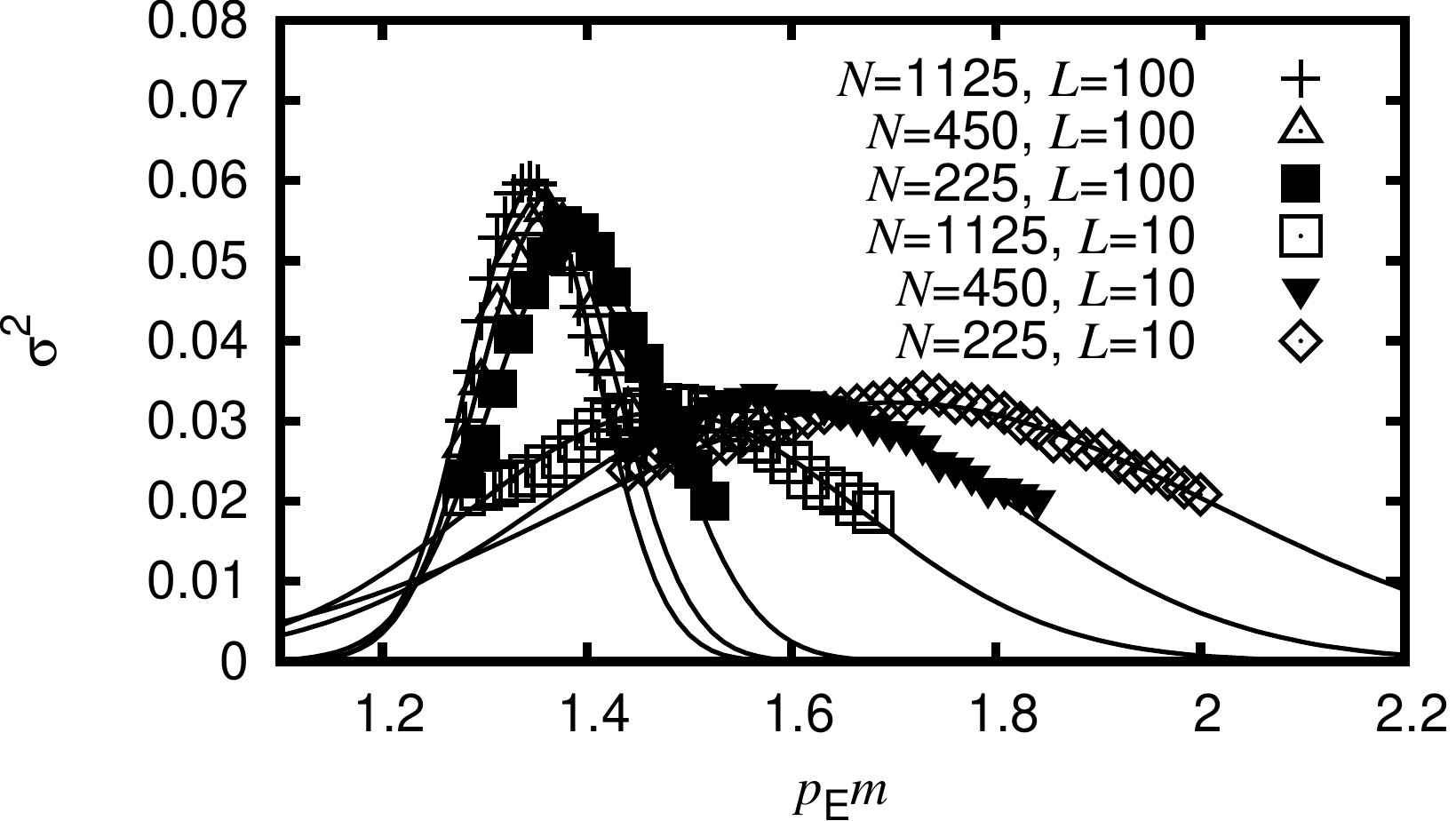}
	\caption{\label{fss_H4}The variance $\sigma^2$ of the total error $E_{\rm tot}$ for the system size $m=n=4$ as function of $p_{\rm E}m$, calculated by using the results of the 1000 HMM realizations.
Data is shown only near the transition regions, respectively.
 The lines show the fits to the Gaussians, which where used to determine the peak positions.}
\end{figure}

As mentioned, a left shift of the decline of $E_{\rm tot}$ is observable 
for all HMM system sizes in Fig. \ref{tot_H4}-\ref{tot_H16} when increasing the learning data 
set sizes. This allows us to determine the phase transition point, which is the position of the 
steepest point of decline in the thermodynamic limit. To determine this we consider the
variances $\sigma^2$ of the total error as a function of $p_{\rm E}m$, which exhibits peaks
at the points of steepest decrease of $E_{\rm tot}$. To obtain estimates for the peak
positions $P_{\rm peak}$ (on the $\tilde p=p_{\rm E}m$ scale) we performed Gaussian fits to the peaks.  Fig. \ref{fss_H4} shows the data used
for the fits and the 
fit results for $n=4$. A shift to the left upon increasing $NL$
is clearly observable. In addition, the Gaussian fits, i.e. the transition regions,
 become narrower for larger 
learning data sets which is often observed in standard finite-size scaling theory 
\cite{cardy1988}. By using the peak positions of all learning data sets, 
we extrapolate the dependence of $P_{\rm peak}(N,L)$ to large learning sets. 
For that, we used the standard finite-size scaling power-law ansatz for
the finite-size dependence of the phase transition positions for second order phase transitions:
\begin{equation}
	P_{\rm peak}(N,L)=\tilde p_{\infty} + a(NL)^{-1/\nu}\,,
	\label{eq:shift_fit}
\end{equation}
where $\tilde p_{\infty}$ denotes the phase transition point in the thermodynamic limit
$NL\to\infty$ and $a$ is a non-universal fit parameter. $\nu$ denotes the exponent 
governing the finite-size corrections and describes in the standard theory of 
continuous phase
transitions the growth of the correlations when approaching the phase transition point

\begin{figure}
	\centering
	\includegraphics[width=0.99\columnwidth]{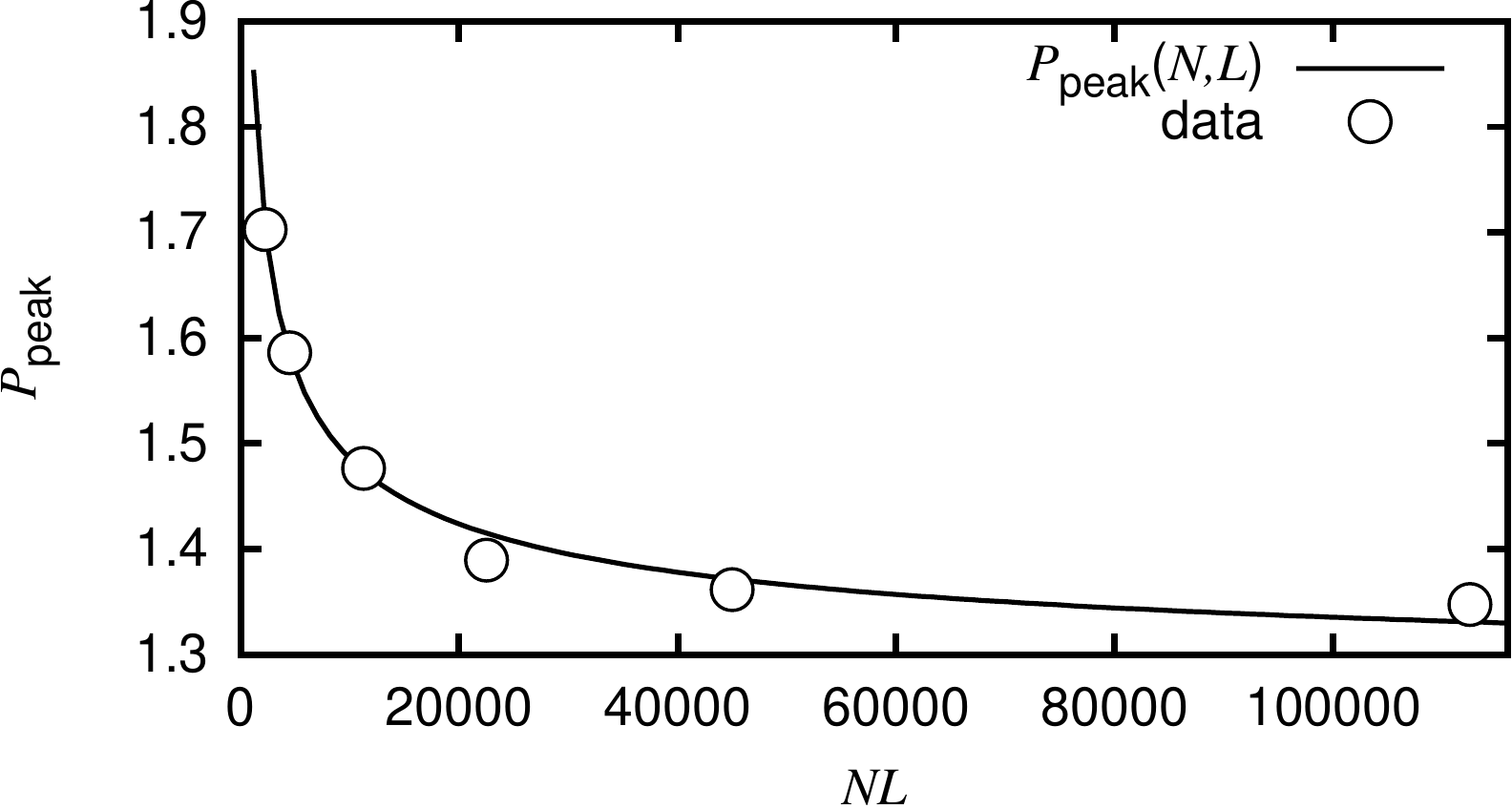}
	\caption{\label{img:shift_pos_H4}Dependence of $P_{\rm peak}$ as a function of the size of learning data set, indicated by the product $NL$. The symbols show the positions of the steepest point of decline which were obtained by the peak positions of the Gaussian fits from Fig.~\ref{fss_H4}.}
\end{figure}

The data for the peak positions together with the fit according to  Eq.~\ref{eq:shift_fit}
for the learning sets of the system size $n=4$ is shown in 
Fig. \ref{img:shift_pos_H4}. On can observe that the fits matches very well, i.e., the
phase transition can be well seen as a continuous phase transition. The resulting fit
parameters, also for the other HMM sizes (which were analyzed in a similar way, not shown
as figures)
are collected in Tab.~\ref{tab:exponents}. We observe that the rescaled critical point
moves to the right with increasing HMM size $n=m$. Also, within (rather large) error bars,
the critical exponents $\nu$ are the same for all HMM sizes, indicating the hard-easy
learning transition is universal with respect to HMM size.

\begin{table}
\begin{tabular}{|r|c|c|}
    \hline
    m & $\tilde p_\infty$ & $\nu$ \\
    \hline
    $4$ & $1.25(4)$ & $2.3(3)$ \\
    $8$ & $1.5(1)$& $2.3(7)$ \\
    $16$ & $2.0(4)$ & $2.1(14)$ \\
    \hline
\end{tabular}
\caption{Rescaled critical points $\tilde p_{\infty}$ (2nd column) and 
critical exponents $\nu$ (4th column) for the different HMM sizes $m$ (1st column)
as obtained from a fit to Eq.~\ref{eq:shift_fit}. For the largest HMM sizes the corrections
to scaling were large such that the peak position for small data size size was omitted
from the fit, resulting in a rather large error bar. 
\label{tab:exponents}}
\end{table}

\begin{figure}
	\centering
	\includegraphics[width=0.99\columnwidth]{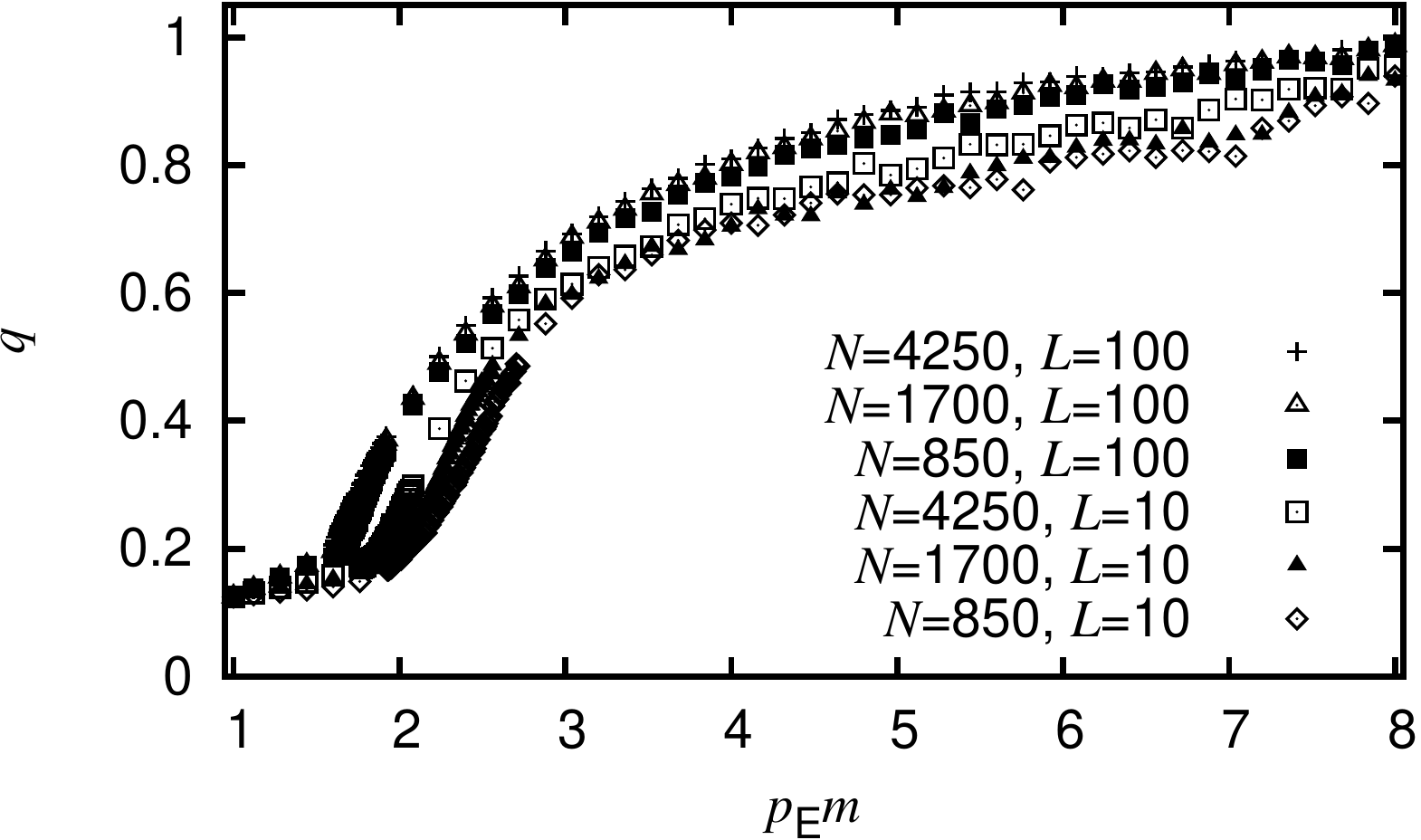}
	\caption{\label{viterbi_H8}The overlap parameter $q$ for HMMs with the size $m=n=8$ and the same learning sets as in Fig.  \ref{tot_H8}.}
\end{figure}

The behavior of the overlap parameter $q$ as a function  of $p_{\rm E}m$ for $n=m=8$ is shown in 
Fig.~\ref{viterbi_H8}. When decreasing the noise, i.e., increasing $p_{\rm E}m$, the
estimated MAP paths become more and more similar to the actual (ground truth) paths.
Since even with the correct parameters estimating the MAP path often does not lead
to the actual path, the behavior is very smooth. Interesting, near the phase transition
point, the curve exhibits a strong kink, which indicates the phase transition is visible
for $q$ as well, but less clearly. 
Note that unlike to the behavior observed previously
\cite{allahverdyan2009}, there is no alternation between several sharp
kinks and monotonously ascents of $q$ over the whole range $p_{\rm E} m \in [1:8]$.
With respect to the finite-size effects of our results, it can be seen that for growing learning
data set size  the transition appears slightly sharper and occurs for smaller
values of $p_{\rm E} m$, i.e., exhibits the same principle finite-size behavior as the total error.
The results of $q$ for the two largest sizes are almost indistinguishable, thus can be taken
to be very similar to the result for the thermodynamic limit $NL\to \infty$.

\begin{figure}
	\centering
	\includegraphics[width=0.99\columnwidth]{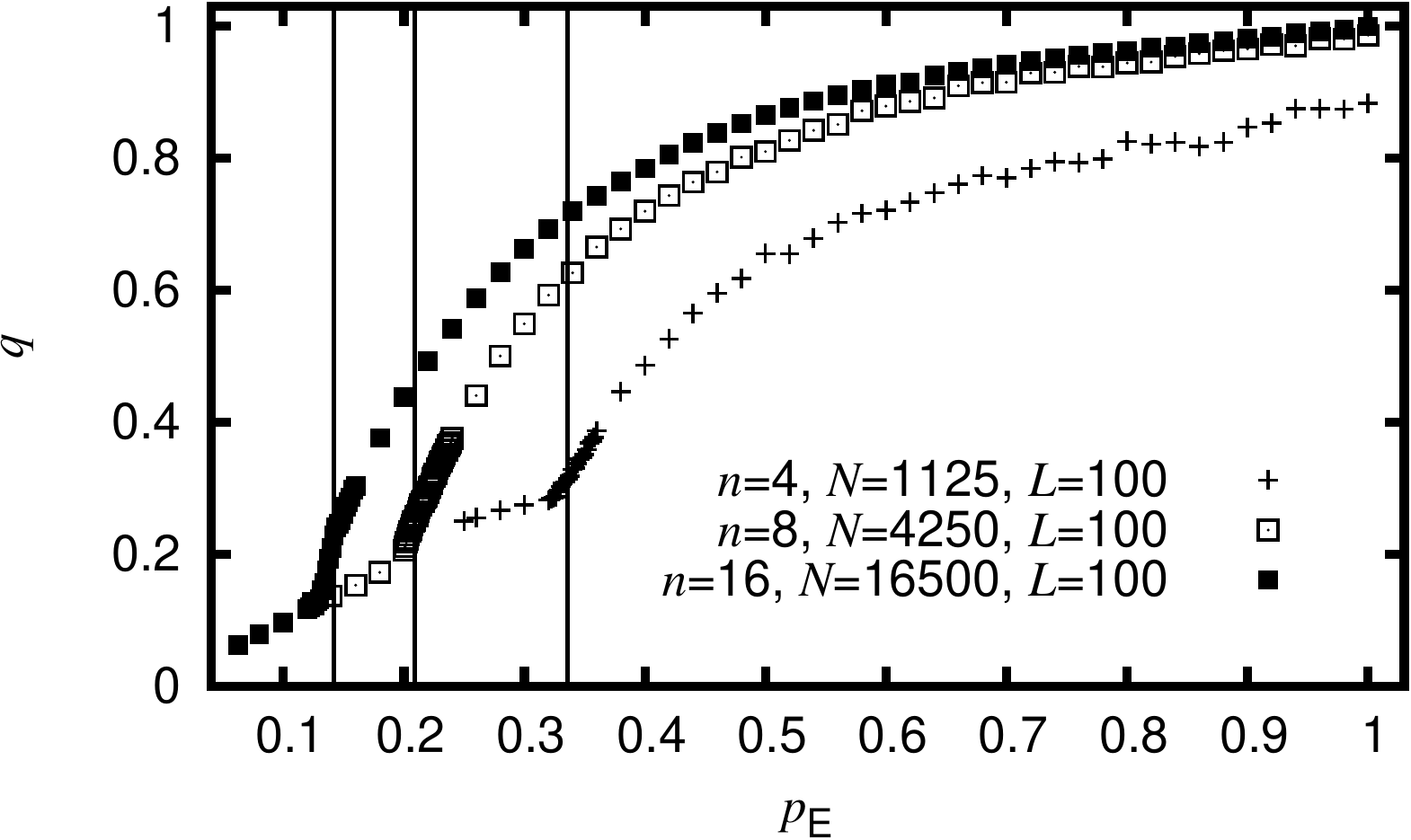}
	\caption{\label{viterbi_all}The overlap parameter $q$ for HMMs with the size 
$n=4; n=8; n=16$ and the largest investigated learning set size as a function of
  the noise parameter $p_{\rm E}$. $p_{\rm E}$ is shown for the
 interval $[\frac{1}{16}:1]$ because its lower bound corresponds to the strongest noise 
level for HMMs with $m=n=16$. The vertical lines indicate the 
extrapolated transition points
obtained from the fit of Eq.~\ref{eq:shift_fit}, and suitably rescaled, i.e., $\tilde p_{\infty}/m$. }
\end{figure}

Thus, to compare the behavior for different HMM sizes, we take always the result obtained
for the largest learning  data set. 
In  Fig. \ref{viterbi_all} a comparison of the overlap parameter $q$ for the different sized HMMs,
is shown here as a function of $p_{\rm E}$ only, because in this way the different curves can be better
distinguished. 
Note that for smallest HMM even the largest learning data set used is
rather small, because this was sufficient to estimate the parameters with high accuracy,
i.e., for a small value of $E_{\rm tot}$. Nevertheless here,
for the overlap, this results in stronger fluctuations as compared to the larger HMMs.
Anyway, one observes that the kinks for $q(p_{\rm E})$ indeed are very close to the
extrapolated  transition points (shown as vertical lines in the figure). Thus,
the hard-to-easy learning phase transition is not only visible in the total error for the
parameters but also in the underlying behavior of the HMMs, as exhibited by the MAP hidden paths.

\section{Summary and Discussion}

In this work we have not used HMMs as tools to analyze physical and other systems, but made
HMMs the subject of interest with a physics perspective, 
like in few previous works, but with a different
research question. 
We have analyzed an ensemble of simple HMMs with $n$ states and $m$ output symbols 
 with respect to learning HMM parameters from
data. We have restricted ourselves to $m=n$.
 For learning we have used the Baum-Welch algorithm to estimate the maximum likelihood parameters.
However, we believe that many aspects of our results also apply for other combinations of $n$ and $m$ and other 
algorithms for parameter estimation.

We have varied a noise parameter $p_{\rm E}$ which controls how much
the visible output symbols convey information about the visited hidden
states. In the limit of $p_{\rm E}\to 1$ no noise exists and perfect
learning is possible, while for $p_{\rm E}\to 1/m$ the output is
completely random and no learning is possible. From analyzing the
error $E_{\rm tot}$  of the learned to the actually used parameters,
from its variance and from the overlap parameter $q$, we obtain clear
evidence for the existence of a non-trivial phase transition between a
``hard'' learning phase and a ``easy'' learning phase.  Note that at
$p_{\rm E}=1/m$ clearly no learning is possible at all. But one could
expect that for any $p_E>1/m$, if the amount of available data is only
large enough, the algorithm could exploit the bias to finally get the
true parameters. For restricting the number of restarts to 10, this is
not the case,  the phase transition point is clearly different from
the trivial limit $1/m$. The transition seems to persist in the limit
of large HMM sizes $n$, since the critical point moves even to the
right on the  $\tilde p_{\rm E} \equiv p_{\rm E}m$ scale with
increasing HMM size $n=m$. Note that it is still possible that
in the ``hard'' phase, the number of local minima is exponential
in the number of states, thus, maybe there is a range of values of $p_{\rm E}$ 
where by using a very large number of restarts one can still find the
ground-truth parameters. To investigate this issue we have,
for $m=8$ and the largest data sets available, 
performed some test runs in the ``hard'' region where
we started the BW algorithm always with the ground-truth parameters.
Indeed, close to the phase transition,
the BW algorithm always stayed close to the ground-truth parameters,
which means that here the phase is only ``hard'' but not ``impossible''.
Nevertheless, close to $p_{\rm E}=1/m$, where the states of the HMM
are indistinguishable, the BW algorithm always iterated away from
the ground-truth parameters, thus, here learning is indeed ``impossible''.

From the finite-size dependence of the
critical points, we have determined the critical exponent $\nu$ for
the different HMM sizes. The value seems to be universal with respect
to HMM size and near $2.3$, but with a rather large error bar.  Thus,
there exist an information-theoretic phase transition in the learning
of the investigated HMMs, similar to transitions observed for neural
networks \cite{seung1992} ,  community detection
\cite{newman2004communityA,ronhovede2010,decelle2011,hu2012} or
optimization algorithms \cite{phase-transitions2005}. Analyzing this
phase transition for HMMs will allow for a better understanding of the
limits of learning.  For example, with more numerical effort, one
could rerun the BW algorithm many times and study the distribution of
local minima and investigate whether they tend to be very close to
each other in parameter space. Or they could turn out to be organized
hierarchically in clusters, similar to the ``replica-symmetry
breaking'' of spin-glasses. Such hierarchical organizations were also
found numerically in the solution landscape of optimization problems
\cite{vccluster2004,sat_cluster2010}.  But such studies about the
behavior of HMMs and parameter learning might also be useful for
practitioners, to optimize algorithms and to get to know meaningful
application ranges.

From a fundamental point of view,
it would be certainly worthwhile to try to use mathematical
(mean-field) methods to analytically perform the disorder average and 
investigate such phase transitions occurring in HMMs
 more thoroughly, expanding 
previous work on two-state HMMs which were tackled by mapping it to Ising 
systems \cite{allahverdyan2009,allahverdyan2015}.

Clearly, we have analyzed only one specific ensemble of HMMs, but we expect 
that many aspects of our results, in particular the existence of one (or more) non-trivial phase
transitions as observed here, holds in general for other types of HMMs. Nevertheless, it would be certainly of interest
to study other HMMs and other types of probabilistic data models in order to study and understand phase transitions of their
learning algorithms. Also it would be worth
investigating different types of parameter-estimation algorithms, 
e.g., to investigate how much the location of the phase transition
depends on the algorithm itself. Nevertheless, due to universality often observed in physical systems,
we expect critical exponents ($\nu$ in our case), describing the growth of correlations
when approaching continuous phase transitions, to also be universal here.
Furthermore, we expect that there are fundamental limits of learning, like
those observed for community detection \cite{decelle2011,hu2012}, where there
is a phase which exhibits statistically significant differences but which 
provably cannot be exploited by any algorithm. 

\bibliography{Literatur}

\end{document}